\documentstyle[11pt]{article}

\renewcommand{\baselinestretch}{1.5}
\renewcommand{\d}{{\rm d}}

\begin{document}

\title{A new method for the calculation of massive multiloop diagrams}
\author{K. Knecht\thanks{%
Email: Kenny.Knecht@rug.ac.be}, H. Verschelde\thanks{%
Email: Henri.Verschelde@rug.ac.be} \\
Department of Mathematical Physics and Astronomy, University of Ghent,\\
Krijgslaan 281, 9000 Ghent, Belgium}
\date{November 1997}
\maketitle

\begin{abstract}
Starting from the parametric representation of a Feynman diagram, we obtain
it's well defined value in dimensional regularisation by changing the
integrals over parameters into contour integrals. That way we eventually
arrive at a representation consisting of well-defined compact integrals. The
result is a simple transformation of the integrand which gives the analytic
continuation of a wide class of Feynmanintegrals. The algorithm will
especially be fit for numerical calculation of general massive multi-loop
integrals. An important advantage of this method is that it allows us to
calculate both infinite {\em and} finite parts independently.
\end{abstract}

\renewcommand{\baselinestretch}{1.5}

\newpage \renewcommand{\baselinestretch}{1.5}

\section{Introduction}

In recent years there has been an increasing interest for the evaluation of
massive multiloop Feynman diagrams. High precision experiments force the
theoretical predictions to reach an equal amount of accuracy. Over the years
several methods have been proposed to deal with the problem.

The first succesful attempts to calculate Feynman-diagrams systematically
were integration by parts \cite{intgparts}, IR-rearrangement method \cite
{IRrearr} and gegenbauer technique \cite{gegenb}. They are not fit though
for calculating finite parts of general massive diagrams, although they have
led to some quite impressive calculations of beta-functions in different
theories \cite{results,chetyrkin}. An excellent review of these methods is
given in \cite{chetyrkin}.

A first general massive approach \cite{lauricella,Davy1,Davy2,Davy3} is
analytic by nature and is based on the following basic principle: by putting
a certain number of masses in the diagram equal to zero a gamma-function
ansatz in obtained, upon which the masses can be added again by means of the
Mellin-Barnes representation. This gives rise to hypergeometric series which
have proven to be rather succesful, e.g. for the asymptotic expansions in
the two loop case \cite{asymt}. These series have certain drawbacks however:
they converge only in certain kinematic regions and although the ansatz
needed for the application of this method can easily be found in the case of
two loops, more loops will give considerable problems. Applications of these
methods are therefore mainly confined to asymptotic 2-loops cases.

Other algorithms avoid the use of Euler-gamma functions and therefore turn
out to be numerical. An excellent example of such a method, which even
avoids parametric representation {\em and} Wick-rotation, relies on the
seperation of an orthogonal space of momenta and integrating out this space
first. Although this method has been succesfully used for some specific
cases \cite{Kreim1,Kreim2,Kreim3,Kreim4}, it has not yet been
expanded up to three loops or more. Other numerical aproaches \cite
{Fuji1,Fuji2} relies on the parametric representation of the diagram. These
approaches also differ mainly form ours in this respect that these methods
are based on a subtraction of divergences under the parametric integral,
while ours is fundamently based on dimensional regularisation. Analytic
continuation in $d$ dimensions is central in our approach: thus we obtain not
only the finite parts but also the poles of the Feynman-diagram. It is
therefore that it has no trouble in dealing with IR and UV divergences at
the same time. Our method is without reservation applicable in a vast number
of cases and easily implemented on a computer.

The rest of the article is organised as follows. Section two contains the
basic formulas of the contour method and some adaptations desirable for
smooth numerical calculations. Section three consists of some examples and
section four is a summary and conclusion.

\section{Contour method}

A scalar diagram $D$ with $L$ loops, $I$ internal lines each labeled by
number $l$ and a mass $m_l$, a total external momentum $P_j$ per vertex and
a space-time dimension $d$ has a parametric representation \cite{Zuber} 
\begin{eqnarray}
I_D(P) = \int_0^\infty \prod_l^I \d\alpha_l \frac{\exp - [ \sum_i \alpha_i
m_i^2 + Q_D(P,\alpha)]} {[(4\pi)^{2L} R_D(\alpha)]^{d/2}}  \label{parameter}
\end{eqnarray}
where 
\begin{eqnarray}
R_D(\alpha) &=& \sum_{{\cal T}} \prod_{l \not{\in }{\cal T}} \alpha_l \\
Q_D(P,\alpha) &=& \frac{1}{R_D(\alpha)}\sum_{{\cal T}^2} s_{{\cal T}^2}
\prod_{l \not{\in }{{\cal T}^2}} \alpha_l
\end{eqnarray}
with ${\cal T}$ the set of all the trees of $D$ and ${{\cal T}^2}$ the set
of all the two-trees of $D$ and $s_{{\cal T}^2}$ the square of the momentum
which passes through the cut. $R_D$ is a homogeneous polynomial in $\alpha$
of degree $L$ and $Q_D$ of degree 1. If the diagram gives rise to tensor
integrals with irreducible numerators we can change these to scalar
integrals using for example the general expression in \cite{recurs2,recurs3}.

The general philosophy of our method will be as follows: we will isolate the
different poles in the integrand and then avoid them by changing the
integral in a contour integral flung around the pole. This way we will
obtain a well-defined analytic continuation of the Feynman integral in
dimensional regularisation.

Our first goal will be to isolate the poles in (\ref{parameter}). Since the
polynomial $R_D$ of e.g. the setting-sun diagram has the form 
\[
\alpha_1 \alpha_2 +\alpha_2 \alpha_3+\alpha_1 \alpha_3 
\]
this is not allways obvious: no simple poles are visible. There is however a
well-known substitution, used in the convergence theorem \cite{Zuber}, which
does the job. It involves a separation of the integration domain into
sectors 
\[
0 \leq \alpha_{\sigma(1)} \leq \alpha_{\sigma(2)} \leq \cdots \leq
\alpha_{\sigma(I)} 
\]
with $\sigma$ a permutation of $(1,2,\ldots,I)$. Per sector we preform the
following change of variables 
\begin{eqnarray}
\alpha_{\sigma(1)} &=& \beta_I \beta_{I-1} \ldots\beta_2 \beta_1  \nonumber
\\
\alpha_{\sigma(2)} &=& \beta_I \beta_{I-1} \ldots \beta_2  \nonumber \\
&\vdots&  \nonumber \\
\alpha_{\sigma(I)} &=& \beta_I  \label{magic}
\end{eqnarray}
where 
\begin{eqnarray*}
0 \leq \beta_I \leq \infty && \\
0 \leq \beta_l \leq 1 && l = 1,\ldots ,I-1
\end{eqnarray*}
Now we can interpret each sector as a family of nested subsets of lines of
the diagram $D$: these nested subsets $S_i \subset S_j$ are determined by
the scaling behaviour of $\beta_i$ and $\beta_j$, $i > j$.

We can prove the following form for $R_D$ \cite{Zuber} 
\begin{equation}
R_D(\beta) = \beta_1^{L_1} \beta_2^{L_2} \ldots \beta_I^{L_I} [1+ {\cal O}%
(\beta) ]  \label{vormR}
\end{equation}
where $L_i$ is the number of independent loops in the subset of lines
specified by the lines which disappear if the corresponding $\beta_i$ is put
to zero (remember: each line corresponds to a certain $\alpha$).
Algebraically we have independent poles in the denominator and now we will
be able to preform our ''contouration''.

The parametric representation (\ref{parameter}) becomes 
\begin{eqnarray*}
I_D(P)& =& \sum_{\sigma} \int_0^\infty \d\beta_I \beta_I^{I-1} \int_0^1
\prod_l^{I-1} \left( \d\beta_l \beta_l^{l-1} \right) \\
&& \times \frac{\exp - [ \sum_i \prod_{j=i}^{I}\beta_j m_{\sigma(i)}^2 +
Q_D^\sigma(P,\beta)]} {[ R_D^\sigma(\beta_1,\ldots,\beta_{I-1})]^{d/2}}
\end{eqnarray*}
thus a sum over permutations $\sigma$, which means a maximum of $I!$ terms.
Thanks to the symmetry in the diagram a certain collecting of terms will
usually be possible. From now on we will omit the sommation over $\sigma$
and concentrate on one sector only. For notational convenience we take $%
\sigma$ to be the identical permutation. Due to homogeneity of $R_D$ and $%
Q_D $ \cite{Zuber} we can immediately preform integration over $\beta_I$ and
get 
\begin{eqnarray}
I_D(P)& =& \frac{\Gamma(I-Ld/2)}{(4\pi)^{2L}}\int_0^1 \prod_l^{I-1} \left(
\d\beta_l \beta_l^{l-1} \right)  \nonumber \\
&& \times \frac{ [ \sum_i \prod_{j=i}^{I-1}\beta_j m_i^2 +
Q_D(P,\beta_1,\ldots,\beta_{I-1})]^{Ld/2-I}} {[
R_D(\beta_1,\ldots,\beta_{I-1})]^{d/2}}  \label{gammaversie}
\end{eqnarray}
We will omit these independent factors in front of the integration in what
follows.

Only $R_D$ gives rise to possible poles if the diagram is globally
UV-divergent (i.e. $L d/2 - I > 0$). If the diagram is UV-convergent and
some masses are zero then poles can come from the numerator in (\ref
{gammaversie}) (which is then in the denominator!): they correspond to IR-divergences. If $L 
d/2 - I > 0$ and
masses are zero, these IR-factors may compensate some UV-poles: this is a well
known fact in dimensional regularisation. Now we can write $I_D$ in the
following form 
\begin{eqnarray}
\lefteqn{I_D(P) = \int_0^1 \d \beta_1 \cdots \int_0^1 \d \beta_{l-1}
\int_0^1 \d \beta_{l+1} \cdots\int_0^1 \d \beta_{I-1}}  \nonumber \\
&& \frac{1}{\beta_1^{p_1} \cdots
\beta_{l-1}^{p_{l-1}}\beta_{l-1}^{p_{l+1}}\cdots \beta_{I-1}^{p_{I-1}}}%
\left( \int_0^1 \d \beta_l \frac{f(\beta)}{\beta_l^{p_l}} \right)  \nonumber
\\
&=&\int_0^1 \d \beta_1 \cdots \int_0^1 \d \beta_{l-1} \int_0^1 \d
\beta_{l+1} \cdots\int_0^1 \d \beta_{I-1}  \nonumber \\
&& \frac{1}{\beta_1^{p_1} \cdots
\beta_{l-1}^{p_{l-1}}\beta_{l-1}^{p_{l+1}}\cdots \beta_{I-1}^{p_{I-1}}}%
I_{(l)}  \label{basis7}
\end{eqnarray}
where the function $f(\beta_1,\ldots,\beta_I)$ is written as as $f(\beta)$
for convenience. We define a unique $p_i$ and $f$ by demanding that $%
f(\beta_i =0) \neq 0$ and that it is analytic at $\beta_i=0$. In dimensional
regularisation, i.e. $d = 4 - 2\varepsilon$, 
\[
p_i = n_i-q_i\varepsilon, 
\]
$n_i $ integer and $q_i$ rational. From now on we will omit the cases $%
n_i\leq 0$ because in these cases the integration can be preformed without
regularisation.

Let's  concentrate on the integration of one specific $\beta_l$. In the
function $f(\beta)$ all other $\beta_i$, $i \neq l$ will be considered to be
parameters for the time being. We will use the notation $f(\beta_l)$. The
function $f(\beta_l)$ is analytic at the origin by definition. In
dimensional regularization (i.e. $q_l \neq 0$) the integrand as a whole is
analytic in a region around the real axis for Re$\beta_l \geq 0$ minus the
cut line along this positive axis. This is true for every $l = 1, \ldots, I$.

We will obtain a meaningful regularized value for the integral $I_{(l)}$ if
we change the integral into a contour-integral, which conincides with the
original integral in non-divergent cases. In concreto we define the $U_l$
(unrenormalized) operator (e.g. \cite{collins}) to be 
\begin{equation}
U_l(I_{(l)}) = \frac{1}{e^{2\pi i p_l}-1}\oint_C \d \beta_l \frac{f(\beta_l)%
}{\beta_l^{p_l}}  \label{prim}
\end{equation}
where the contour $C$, as is showed in figure 1 
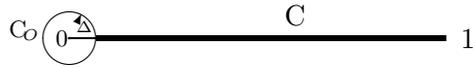
\begin{figure}[tbp]
\unitlength=1mm \special{em:linewidth 0.4pt} \linethickness{0.4pt} 
\begin{picture}(110.00,40.00)
\put(30.00,22.00){\circle{7.5}}
\put (30.00,22.00){\line(1,0){50.00}}
\put (33.65,22.25){\line(1,0){46.35}}
\put (33.65,21.75){\line(1,0){46.35}}
\put(29.00,22.00){\makebox(0,0)[cc]{\footnotesize 0}}
\put(30.00,22.00){\circle*{0.00}}
\put(30.00,80.00){\circle*{0.00}}
\put(60.00,25.00){\makebox(0,0)[cc]{C}}
\put(83.00,22.00){\makebox(0,0)[cc]{1}}
\put(24.00,23.00){\makebox(0,0)[cc]{{\footnotesize C$_{\! O}$}}}
\put(30,22){\vector(1,2){1.65}}
\put(32.00,23.50){\makebox(0,0)[cc]{{\tiny $\Delta$}}}
\end{picture}
\caption{The contour for obtaining analytic continuation of $\beta$%
-integrals }
\end{figure}
, indeed avoids the pole $\beta_l=0$ by a tiny contour $C_0$ which is a
circle with radius $\Delta$, $\lim \Delta \rightarrow 0+$ and the integrand
is analytic in the region of the contour as shown before. It is clear that
this definition only holds for non-integer $p_l$: our method is intimately
connected to dimensional regularization.

In order to get rid of contours again, we preform integration over $C_0$
explicitly. We are abel to do so if we write $f(\beta_l)$ in a Taylor
expansion (this is allowed due to analyticity of $f$ at the origin). Since $%
\lim \Delta \rightarrow 0+$ we keep the $n_l$ first terms of the Taylor
expansion. Other terms vanish in this limit, they are namely ${\cal O}%
(\Delta^{p_l\varepsilon})$ and higher. Finally we obtain 
\[
U_l(I_{(l)}) = \int_\Delta^1 \d \beta_l \frac{f(\beta_l)}{\beta_l^{p_l}}
+\sum_{j=0}^{n_l-1} \frac{\Delta^{j-p_l + 1 }}{j! (j-p_l+1) } \left. \frac{%
\partial^j f(z_l)}{\partial^j z_l} \right|_{z_l=0} 
\]
This form still contains $\Delta$ explicitly. If we use the identity for
every $p_l \neq 0$ 
\begin{eqnarray*}
\int_\Delta^1 \d \beta_l \: \beta_l^{p_l-1} &=& \frac{1}{p_l} - \frac{%
\Delta^{p_l} }{p_l}
\end{eqnarray*}
and regroup the terms, we find for the case $n_l \geq 1$ 
\begin{eqnarray*}
U_l(I_{(l)}) &=& \int_\Delta^1 \d \beta_l \left( f(\beta_l)\beta_l^{-p_l} -
\sum_{j=0}^{n_l-1} \frac{\beta_l^{j-p_l} }{j!} \left. \frac{\partial^j f(z_l)%
}{\partial^j z_l} \right|_{z_l=0} \right) \\
&&+ \sum_{j=0}^{n_l-1} \frac{1 }{j! } \left( \frac{1}{j-p_l+1} \right)
\left. \frac{\partial^j f(z_l)}{\partial^j z_l} \right|_{z_l=0}
\end{eqnarray*}
Noticing that the integrand under the first integral {\em as a whole} is of
the order $\beta_l^{q_l\varepsilon}$, the integral will be convergent in the
limit $\Delta \rightarrow 0$. We will use the following notation 
\[
f_{(l)}^{(*)} = \left( f(\beta_l)\beta_l^{-p_l} - \sum_{j=0}^{n_l-1} \frac{%
\beta_l^{j-p_l} }{j!} \left. \frac{\partial^j f(z_l)}{\partial^j z_l}
\right|_{z_l=0} \right) 
\]
and for the other terms 
\[
f_{(l)}^{(j)} = \frac{1 }{j! } \left( \frac{1}{j-p_l+1} \right) \left. \frac{%
\partial^j f(z_l)}{\partial^j z_l} \right|_{z_l=0} 
\]
They are independent of $\beta_l$.

We can write this result as a simple transformation of the integral 
\begin{eqnarray}
f_{(l)} &=& U_l(I_{(l)})  \nonumber \\
&=&U_l(\int_0^1 \d \beta_l \frac{f(\beta_l)}{\beta_l^{p_l}})  \nonumber \\
&= &\int_0^1 \d \beta_l f_{(l)}^{(*)} + \sum_j f_{(l)}^{(j)}
\label{hetresult}
\end{eqnarray}
The next difficulty we have to face, is whether or not this operation can be
repeated for other $\beta$'s? Potential problems arise when $\sum_i \beta_i
m_i^2 + Q_D(P,\beta)$ comes in the denominator: we have not been able to
prove an expression like (\ref{vormR}) for it, but this appears to be always
the case. We have done numerous tests on diagrams up to three loops under
different circumstances (several external momenta, masses or massless,...)
and have found no counter-examples. Yet a rigorous mathematical proof that
this applies for {\em every} diagram is still lacking. Here we will assume
that the diagrams we are dealing cause no problems.

In that case if we have carried out the $U_{l_1}$-operation, we can
re-establish the other $\beta_i$ as true variables. Then we choose another $%
l_2$ to focus on, write 
\[
I_{(l_1,l_2)} =\int_0^1 \d \beta_{l_2 } \frac{f_{(l_1)}}{%
\beta_{l_2}^{p_{l_2}}} 
\]
and apply the operator $U_{l_2}$ to $I_{(l_1,l_2)}$. We can maintain our
notation by adding a number to the sub- and superscript of $f$, e.g. after $%
U_2 \circ U_1$ we will get 
\[
f_{(l_1,l_2)} = f_{(l_1,l_2)}^{(*,*)} + \sum_i f_{(l_1,l_2)}^{(i,*)} +
\sum_j f_{(l_1,l_2)}^{(*,j)} + \sum_{i,j} f_{(l_1,l_2)}^{(i,j)} 
\]
We repeat this procedure until no $\beta$-poles are left, thus regulazing
the feynman-integral $I_D(P)$ completely.

Several remarks are in order.

\begin{itemize}
\item  We can regulaze every variable $\beta_i$ before preforming
integration, i.e. $U_{(i)}$ commutes with $\int \d \beta_j$ for $i \neq j$. We
only have to take care to keep the different parts of $f_{(l)}^{(*)}$ (all
other indices are arbitrary: $j$ or $*$) together, because each term on his
own has a pole in $\beta_l$, only the sum converges. Before preforming
integration we can also expand in $\varepsilon$. This allows us to {\em %
calculate the different coefficients of the laurent-expansion in $%
\varepsilon $ seperately}. Thus the $T$ operation of \cite{collins}
(isolating the divergent part) is easily implemented. Higher pole parts can
only come from the $f_{(l)}^{(i)}$ parts.

\item  The result of the different operations $U_{l_1} \circ U_{l_2} \circ
\ldots \circ U_{l_n}$ is independent of their order. If there are only poles
in $\beta_1$ and $\beta_2$, so $U_{l_1}\circ U_{l_2}$ completely regularizes
the integral then: 
\[
U_{l_1}\circ U_{l_2} = U_{l_2}\circ U_{l_1} 
\]
This follows from the fact that the $U$-operators are in fact analytic
continuations (see our primary definition (\ref{prim})): if ${\rm Re} p_l
\leq 1$ the contour-integrals coincide with the ordinary integrals which are
convergent, so $U_{l_1}\circ U_{l_2} = U_{l_2}\circ U_{l_1}$ in the area $%
{\rm Re} p_l \leq 1$. By the principle of analytic continuation they must
coincide for all complex $d$ (which fixes all the $p_i$'s), for which $%
U_{l_1}$ and $U_{l_2}$ exist. Of course this implies that for every variable 
$\beta_i$ this analytic continuation has been carried through.
\end{itemize}

It is clear that an expression like (\ref{hetresult}) will be especially
useful in numerical calculations and therefore it might be useful to examen
it's numerical behaviour. Only $f_{(l)}^{(*)}$ will cause trouble around
zero. Although the function as a whole is finite, it is really a subtraction
of one or more infinite values. Moreover it is of the order $%
\beta_l^{q_l\varepsilon}$ which after expanding in $\varepsilon$ gives $1+
q_l\varepsilon \ln(\beta_l)+ \cdots$ and is thus another thread to numerical
stability. Therefore we will approximate $f_{(l)}^{(*)} $ around zero, by
the next term in the Taylor expansion, i.e. 
\[
f_{(l)}^{(*)} \approx \left. \frac{\partial^{n_l} f(z_l)}{\partial^{n_l} z_l}
\right|_{z_l=0} \frac{\beta_l^{q_l \varepsilon}}{n_l!} 
\]
Introducing this approximation in the interval $[0,\delta]$, $\delta \ll 1$
being a certain numerical value, we obtain 
\begin{equation}
\int^1_0 \d \beta_l\frac{f_{(l)}^{(*)} }{\beta_l^{p_l}} \approx \left. \frac{%
\partial^{n_l} f(z_l)}{\partial^{n_l} z_l} \right|_{z_l=0} \frac{%
\delta^{1-q_l\varepsilon}}{(1-q_l\varepsilon) n_l!} + \int^1_\delta \d
\beta_l\frac{f_{(l)}^{(*)}}{\beta_l^{p_l}}  \label{numredding}
\end{equation}
The intrinsic error of this approximation depends on the highest order pole.
If it is $1/\varepsilon^L$, then terms proportional to $\ln(\delta)^{L-1}$
will appear. If $\delta$ is very small then this correction will be large,
resulting in a smaller accuracy. Thus there is a best value for $\delta$
typical for every diagram. Typical values are e.g. $\delta = 10^{-5}$. In
order to obtain larger values for $\delta$ without losing accuracy one could
include more terms of the taylor expansion.

\section{Examples}

In this section we will discuss one example at length: the famous
''setting-sun'' diagram in scalar $\lambda\phi^4$. Then some other results
for diagrams with a larger number of propagators are included. In order to
be able to compare with analytic results, we begin by choosing some trivial
vacuumbubbles: no external momenta and equal masses. Of course this is only
to make comparison possible: it is the analytic methods that are bound by
these limitations, not the contour method as will be shown in some explicit
numerical examples.

We put the momentum $p$ equal to zero and choose the masses equal. This
makes the diagram is very symmetrical: topologically every leg is equivalent
and this will simplifie the explicit calculation.

Starting from the well-known parametric representation of the setting-sun
diagram, we get only one $\beta$-sector (all masses are equal). Applying
formula (\ref{gammaversie})), we get 
\begin{equation}
I_{\mbox{ss}} =\frac{(m^2)^{d-3}}{(4\pi)^{d}} \Gamma(3-d)3! \int_0^1
\d\beta_1 \int_0^1 \d\beta_2 \frac{(1+\beta_2 + \beta_1\beta_2)^{d-3}} {%
(1+\beta_1 + \beta_1\beta_2)^{d/2}} \frac{1}{\beta_2^{d/2-1}}
\end{equation}
The diagram is globally divergent in $d \geq 3$ whence the factor $%
\Gamma(3-d)$. At this stage we see a simple pole in $\beta_2$ emerging. This
corresponds to the two-legged subdivergence present in the diagram
(renormalization of the coupling constant). With the notations of section 2,
we have 
\[
f(\beta_2) = \frac{(1+\beta_2 + \beta_1\beta_2)^{d-3}} {(1+\beta_1 +
\beta_1\beta_2)^{d/2}} 
\]
and $p_2= n_2-q_2 \varepsilon= 1-\varepsilon$. Applying formula (\ref
{hetresult}) we get 
\begin{eqnarray*}
\lefteqn{U_{(2)} \left( \int_0^1 \d\beta_2 \frac{(1+\beta_2 +
\beta_1\beta_2)^{d-3}} {(1+\beta_1 + \beta_1\beta_2)^{d/2}} \frac{1}{%
\beta_2^{d/2-1}} \right) } \\
&=& \int_0^1 \d\beta_2 \left( \frac{(1+\beta_2 + \beta_1\beta_2)^{d-3}} {%
(1+\beta_1 + \beta_1\beta_2)^{d/2}} - \frac{1}{(1+\beta_1)^{d/2}}\right)%
\frac{1}{\beta_2^{d/2-1}} \\
&& + \frac{1}{(2-d/2)(1+\beta_1)^{d/2}}
\end{eqnarray*}
These integrals are finite, as expected. If we restrict ourselves to the
divergent part (without expanding the $\Gamma(3-d)$ though, thus $%
f_{(2)}^{(*)}$ and the $f_{(2)}^{(j)}$'s up to $\varepsilon^0$) 
we get 
\begin{eqnarray*}
f_{(2)}^{(*)} &=& \int_0^1 \d\beta_2 \left(\frac{1+\beta_2 + \beta_1\beta_2%
} {(1+\beta_1 + \beta_1\beta_2)^2} - \frac{1}{(1+\beta_1 )^2} \right) \frac{1%
}{\beta_2^1} \\
f_{(2)}^{(0)}& =& \frac{1}{\varepsilon(1+\beta_1)^2} + \frac{\ln(1+\beta_1)%
} {(1+\beta_1)^2}
\end{eqnarray*}
We can preform this integration analytically 
\begin{eqnarray*}
\int_0^1 \d \beta_2 \int_0^1 \d \beta_1 f_{(2)}^{(*)} &=& \frac{\ln 2}{2} \\
\int_0^1 \d \beta_1 f_{(2)}^{(0)}& =& \frac{1}{2\varepsilon} + \frac{1- \ln 2%
}{2}
\end{eqnarray*}
If we expand further in $\varepsilon$ and preform the numerical calculations
we obtain the finite part as well: 
\begin{equation}
I_{\mbox{ss}} =\frac{(m^2)^{d-3}}{(4\pi)^{d}} \Gamma(3-d) \left(\frac{3}{
\varepsilon} + 3 - 8.966523919 \varepsilon \right)
\end{equation}
Comparing this result by numerical evaluation of the analytic results in 
\cite{Davy1} we see that only the last digit is different.

Another test we did was to repeat some of the cases from \cite{lauricella},
i.e. a subtracted setting-sun diagram with arbitrary masses and external
momentum. A selection \newline
\newline
\newline
\begin{tabular}{|r|r|r|r|r|r|}
\hline
$p^2$ & $m_1$ & $m_2$ & $m_3$ & Contour & Series \\ \hline
9 & 3 & 3 & 10 & -7.31299 & -7.31298 \\ 
49 & 1 & 1 & 10 & -0.316742 & -0.31675 \\ 
-25 & 2 & 2 & 10 & -1.942844 & -1.94285 \\ 
-250 & 4 & 4 & 4 & -13.57190 & -13.5719 \\ 
100 & 3 & 3 & 3 & -1.282852-$i$20.89960 & -1.28285-$i$ 20.8996 \\ 
49 & 20 & 20 & 10 & -1014.695748 &  \\ \hline
\end{tabular}
\newline
\newline
\newline
The last result cannot be calculated with the method of \cite{lauricella}
because $m_1 + m_2 < m_3$ should apply under the threshold.

Another test for our method was the so-called ''basketball-diagram'', a
vacu\"umdiagram in $\lambda\phi^4$ with 3 loops and 4 legs. We also assumed
equal masses here in order to be able to compare to known analytic results.
We have calculated the diagram in four cases: with one, two, three and four
massive legs.

The numerical procedure as given by formula (\ref{numredding}) in section
two, was implemented in MATHEMATICA \cite{math}: the diagram, expressed in
it's original form is automatically transformed into the parametric $\alpha$%
-form and $\beta$-form, then cast into the expression (\ref{numredding}) and
numerically evaluated. Because we are dealing with a very limited class of
functions, it is possible to program a package that preforms one integration
analytically. This general procedure was used to evaluate all the following
results.

These were the results we obtained up to order $\varepsilon^0$. We compare
these values to analytic results in \cite{analytic}. All results are given
up to a factor $\frac{(m^2)^{4-3d/2}}{(4\pi)^{3d/2}} \Gamma(4-3d/2)$.
Numerical evaluations of the analytic results are only shown in 
relevant cases. \newline
\newline
\begin{tabular}{|l||r|r|r|r||r|r|}
\hline
Type & \multicolumn{4}{|c|}{Contour} & \multicolumn{2}{|c|}{Analytic} \\ 
\hline
& $\varepsilon^{-2}$ & $\varepsilon^{-1}$ & $\varepsilon^{0}$ & $%
\varepsilon^{1}$ & $\varepsilon^{0}$ & $\varepsilon^{1}$ \\ \hline
1 mass &  & -0.5 & -1.5 & -5.967401126 & -1.5 & -5.967401100 \\ 
2 masses & 2 & -2 & -19.86960443 & 22.33542372 & -19.86960440 & 22.335425302
\\ 
3 masses & 6 & - 4.5 & -55.10881324 & 44.04808956 & -55.10881320 & 
44.04808897 \\ 
4 masses & 12 & -8 & -107.2176265 & 26.87588164 & -107.2176264 & 26.87588031
\\ \hline
\end{tabular}
\newline
\newline
These results were obtained with $\delta = 10^{-5}$ and relative errors are
of order $10^{-8}$.

Other diagrams we evalutated are the 3-loop, 5 propagator diagram with 2
masses 
\[
I_5 = \int \d k \frac{1}{(k_1 - k_3)^2(k_1 - k_2)^2(k_2 -
k_3)^2(k_3^2+m^2)(k_2^2+m^2)} 
\]
and the mercedes diagram with 3 masses in a very asymetric configuration 
\[
I_6 = \int \d k \frac{1}{k_1^2 (k_2^2+m^2) k_3^2((k_1 - k_3)^2+m^2)(k_1 -
k_2)^2((k_2 - k_3)^2+m^2)} 
\]
where $\d k$ stands for integration over all internal momenta $k_i$. As a
result we found: 
\[
I_5 = \frac{(m^2)^{5-3d/2} \Gamma(5-3d/2)}{(4 \pi)^{3d/2}} \left[ \frac{1}{%
\varepsilon^2}+ \frac{3}{\varepsilon} + 2.065197729 + 19.42850404
\varepsilon\right] 
\]
and 
\[
I_6 = \frac{(m^2)^{6-3d/2} \Gamma(6-3d/2)}{(4 \pi)^{3d/2}} \left[ 
7.21234140 -9.081028480 \varepsilon\right] 
\]
In the case of $I_5$ we encounter relative errors of order $10^{-8}$ in
comparison with analytic results. According to \cite{analytic} $I_6$ has not
yet been evaluated analytically. Our numerical results differ only by order $%
10^{-9}$ from theirs. We can conclude that the numerical behaviour of the
method is fine.

\section{Summary and conclusion}

The main merit of our contour-method as introduced here, lies evidently in
its general character. Indeed it is applicable for a wide class of
Feynman-diagrams scalar {\em and} tensorial (although the borders of this
class or not yet quite clear). 
The computer-algorithm based on expression (\ref{numredding}) is easy to
implement and virtually independent of the actual diagram fed to the system.

In the section examples we confined ourselves mainly to diagrams for which
analytic expressions are known, in order to investigate numerical behaviour.
As shown in the case of the setting-sun diagram our contour-method is
without further trouble applicable in the cases of external momenta and
abitary masses and that is where her strength lies.

As remarked our contour-method is a numerical method in the first place. The
necessary calculations involve especially a lot of numerical integrations,
which are known not to be an easy problem. Even at this point our method has
some advantages: the integration intervals are compact at all time (\ref
{hetresult}) and -thanks to our numerical adjustment (\ref{numredding})- all
integrands are finite. Such calculations can best be done by some adaptive
Monte-Carlo method. The method will also give rise to a large number of
integrals, but this drawback can partially be met using the recursion
algoritms in \cite{recurs1,recurs2, recurs3}, forcing us only to calculate a
small number of so-called master-integrals. This is a common practice: such
techniques have to be used in any method involving higher-order
Feynman-diagrams in order to reduce the number of calculations.

To sum up, we have a method which is easy the implement and applicable in a
vast number of cases, even those where tradional methods fail, such as
finite parts of complicated diagrams. Furthermore we are able to calculate
the different coefficients of the laurent-expansion in $\varepsilon $
independently. Therefore it is a useful tool for high-precision calculation
of general massive Feynman-diagrams.

\end{document}